\documentclass[aps,prd,amsmath,amssymb,preprintnumbers,preprint,nofootinbib,11pt]{article}
\usepackage{epsfig}
\usepackage{colortbl}
\usepackage{amsthm}
\usepackage{graphics}
\usepackage{graphicx}
\def\be{\begin{equation}}
\def\ee{\end{equation}}
\def\bea{\begin{eqnarray}}
\def\eea{\end{eqnarray}}
\usepackage{graphicx}
\usepackage{pxfonts}
\usepackage{epstopdf}
\usepackage{floatpag}
 \usepackage{colortbl}
 \usepackage{youngtab}
\usepackage{slashed}
\definecolor{rossoCP3}{cmyk}{0,.88,.77,.40}

\renewcommand{\(}{\left(}
\renewcommand{\)}{\right)}

\def\beq{\begin{equation}}
\def\eeq{\end{equation}}
\def\be{\begin{equation}}
\def\ee{\end{equation}}
\def\bea{\begin{eqnarray}}
\def\eea{\end{eqnarray}}
\def\ba{\begin{eqnarray}}
\def\ea{\end{eqnarray}}

\definecolor{red}{rgb}{1.00,0.00,0.00}

\catcode`\@=11
\def\lsim{\mathrel{\mathpalette\@versim<}}
\def\gsim{\mathrel{\mathpalette\@versim>}}
\def\@versim#1#2{\vcenter{\offinterlineskip
\ialign{$\m@th#1\hfil##\hfil$\crcr#2\crcr\sim\crcr } }}
\catcode`\@=12

\catcode`@=12
\begin{document}
\thispagestyle{empty}
\begin{flushright}
July 14,  2014\
\end{flushright}
\vspace{0.3in}
\begin{center}
{\LARGE \bf {\color{rossoCP3} BICEP2 constrains Composite Inflation}\\}
\vspace{1.2in}
{\Large Phongpichit Channuie$^1$\\}
\vspace{0.2in}
{\sl $^1$ Department of Physics, School of Science, Walailak University, \\Thasala District, Nakhon Si Thammarat, 80160, Thailand\\}
\vspace{0.2in}
{Email: phongpichit.ch@wu.ac.th\\} 
\end{center}
\vspace{0.6in}
\begin{abstract}\
In light of BICEP2, we reexamine single field inflationary models in which the inflaton is a composite state stemming from various four-dimensional strongly coupled theories. We study in the Einstein frame a set of cosmological parameters, the primordial spectral index $n_{s}$ and tensor-to-scalar ratio $r$, predicted by such models. We confront the predicted results with the joint Planck data, and with the recent BICEP2 data. We constrain the number of e-foldings for composite models of inflation in order to obtain a successful inflation. We find the minimal composite inflationary model is fully consistent with the Planck data. However it is in tension with the recent BICEP2 data. The observables predicted by the glueball inflationary model can be consistent with both Planck and BICEP2 contours if a suitable number of e-foldings are chosen. Surprisingly, the super Yang-Mills inflationary prediction is significantly consistent with the Planck and BICEP2 observations.
\\\\\\\
{PACS: 1.10.-z,\,11.30.Rd,\,98.80.-k\\} 
{Keywords: Composite Inflation,\,Non-minimal Coupling,\,Strongly Interacting Field Theories,\\ {\sc Planck} and {\sc Bicep2} Constraints.} 
\end{abstract}

\newpage
\tableofcontents

\section{Introduction}

Nowadays, inflationary models gain a lot of interest. The inflationary paradigms \cite{Alex,KSa,KSa1,DKa,GUT} tend to solve important issues, e.g. the magnetic monopoles, the flatness, and the horizon problems, and subsequently provide the mechanism for generation of density perturbations as seed for the formation of large scale structure in the universe. Most models of inflation so far were formulated by introducing new scalar fields (called inflaton) with a nearly flat potential appearing in many paradigms, e.g.~\cite{new,new1,chaotic,natural,natural1}, and even in superstring \cite{Espinosa:1998ks,Casas:1999xj} and supergravity theories \cite{Yamaguchi:2011kg,Farakos:2013cqa,Linde:2013aya,Ferrara:2013rsa}.

However, the theories featuring elementary scalar fields are unnatural meaning that quantum corrections generate unprotected quadratic divergences which must be fine-tuned away if the models must be true till the Planck energy scale. Hence, this is the main reason why the authors in \cite{Channuie:2011rq,Bezrukov:2011mv,Channuie:2012bv} investigated inflation in which the inflaton need not be an elementary degree of freedom. Recent investigations show that it is possible to construct models in which the inflaton emerges as a composite state of a four-dimensional strongly coupled theory. We called these alternative paradigms composite inflation. There were other models of super or holographic composite inflation \cite{Cvetic:1989eg,Thomas:1995dq,GarciaBellido:1997mq,Allahverdi:2006iq,Hamaguchi:2008uy,Evans:2010tf,Alberghi:2010sn,Alberghi:2009kk}. In order to constrain the inflationary theory, recent investigations contain hundreds of different scenarios \cite{Martin:2013tda}.

Most frequently, we added a term in which the inflaton field, $\phi$, non-minimally coupled to gravity to the action, say $\xi\phi^{2}R/M^{2}_{\rm P}$. This term has purely phenomenological origin. The reason resides from the fact that one want to relax the unacceptable large amplitude of primordial power spectrum generated if one takes $\xi=0$ or small. However, 
it is instructive here to provide some insight on how a non-minimal coupling can be naturally formed. An instructive analysis of the generated coupling of a composite scalar field to gravity has been initiated in the Nambu-Jona-Lasinio (NJL) model \cite{Hill:1991jc}. In this case, a composite field is a chiral condensate, $H$, from the NJL model which couples to gravity via a coupling term very similar to that we deploy in this work. The presence of the induced non-minimal coupling of the boundstate object to gravity, $\xi H^{\dagger}HR$ with $\xi$ a coupling constant, and $R$ the (Ricci) scalar curvature, can be implemented from the symmetric phase of the theory, i.e. a massless case, or from the broken phase when $\textless H\textgreater=v\neq0$ providing the non-minimal term, $\sqrt{2}\xi v \phi R$ where $H\equiv v+\phi/\sqrt{2}$ with $v$ the vacuum expectation value of $H$. However, the results of $\xi$ are the same. In principle, we can transform such a term into another form by applying the conformal transformation (see detailed discussion in \cite{FuMa}). Among many frames, the Jordan frame and the Einstein frame are those discussed in the community. Roughly speaking, physics looks different in two different conformal frames. However, physical conclusions remains the same in a weak gravitational field limit \cite{FuMa}. 

In this work, we study observables, i.e. the scalar spectral index $n_{s}$ and tensor-to-scalar ratio $r$, in the Einstein frame predicted by the composite inflationary models recently proposed. We will briefly present the concept of conformal transformation. We then compute the power spectrum for the primordial curvature perturbations and the tensor-to-scalar ratio for various composite inflationary models in the Einstein frame. Next we constrain the models by placing our results into the $(r, n_{s})$ plane implemented by using the observational bound for $n_s$ and $r$ from the Planck data and also confront the results with the recent BICEP2 data. Finally, some comments about our results are made in the last section.

\section{Composite Setup and Conformal Transformation}
\label{action}

We consider a generic strongly-interacting theory non-minimally coupled to gravity. For the model of inflation, we identify the inflaton with one of the lightest composite states of the theory and denote it with $\Phi$. This state has mass dimension $d$.  The action for the composite models studied below can be written in the general form 
\begin{eqnarray}
\mathcal{S}_{\rm J}=\int d^{4}x \sqrt{-g}\left[\frac{1}{2}{\cal F}(\Phi)R-{\cal G}(\Phi)g^{\mu\nu}\partial_{\mu}\Phi\partial_{\nu}\Phi-{\cal V}({\Phi})\right]\,, \label{nonminimal}
\end{eqnarray}
where ${\cal F}(\Phi)$ and ${\cal G}(\Phi)$ can in general be an arbitrary function of the composite field $\Phi$. ${\cal F}(\Phi)$ gives rise to the non-minimal coupling $\xi\Phi^{2/d} R$. To recover the ordinary Einstein theory of gravity at low energy, we deduce ${\cal F}(v)=M^{2}_{P}$ where $v$ is the vacuum expectation value of the field $\Phi$ at the end of inflation and $M^{2}_{P}=(8\pi G)^{-1/2}\simeq 2.436\times 10^{18}\,{\rm GeV}$ is the reduced Planck energy scale. According to the models of single-field inflation \cite{Channuie:2011rq,Bezrukov:2011mv,Channuie:2012bv} in which the inflaton is a composite state stemming from various strongly-interacting theories well-known in particle physics, in this work, we write ${\cal F}(\Phi)$ and ${\cal G}(\Phi)$ in the form as follows:
\begin{equation}
{\cal F}(\Phi) = M^{2}_{\rm P}+\xi\Phi^{2/d}\,,\quad {\cal G}(\Phi) = \Phi^{(2-2d)/d}\,,
\end{equation}
where the non-minimal coupling to gravity is controlled by the dimensionless coupling $\xi$. With the non-minimal coupling term phenomenologically introduced, one need to diagonalize into another form by applying a conformal transformation. The conformal transformation transforms a metric tensor $g_{\mu\nu}$ into another metric $\tilde{g}_{\mu\nu}$ according to the rule
\begin{eqnarray}
g_{\mu\nu}\rightarrow\tilde{g}_{\mu\nu}=\Omega^{2}(\Phi)g_{\mu\nu}\quad\quad{\rm with}\quad\quad\Omega^{2}(\Phi)=\frac{{\cal F}(\Phi)}{M^{2}_{P}}\,, \label{scale}
\end{eqnarray}
such that 
\begin{eqnarray}
\tilde{g}^{\mu\nu}=\Omega^{-2}(\Phi)g^{\mu\nu}\quad\quad{\rm and}\quad\quad\sqrt{-\tilde{g}}=\Omega^{-4}(\Phi)\sqrt{-{g}}\,.\label{scale2}
\end{eqnarray}
Hereafter we drop tildes to express variables in the Einstein frames. Applying the conformal transformation leads to the Einstein frame and the action reads
\begin{eqnarray}
\mathcal{S}_{\rm CI,E} &=&\int d^{4}x \sqrt{-g}\Big[\frac{1}{2} M_{\rm P}^2 R - \left(\frac{M_{\rm P}^2{\cal G}(\Phi)}{{\cal F}(\Phi)} + \frac{3}{2} M^{2}_{\rm P} \left(\frac{{\cal F}(\Phi)'}{{\cal F}(\Phi)}\right)^{2} \right)g^{\mu \nu} \partial_{\mu} \Phi \partial_{\nu} \Phi  \nonumber\\&&\quad\quad\quad\quad\quad\,\,- \Omega ^{-4} V({\Phi}) \Big]\,, \label{Einsteinframe}
\end{eqnarray}
where prime denotes derivative over the field $\Phi$. We arrived at an involved kinetic term for the inflaton. It is convenient to introduce a canonically normalized field $\chi$ related to €$\Phi$ via
\begin{eqnarray}
\frac{d\chi}{d\Phi}=\sqrt{2}M_{\rm P}\sqrt{\frac{{\cal G}(\Phi)}{{\cal F}(\Phi)} + \frac{3}{2}\left(\frac{{\cal F}(\Phi)'}{{\cal F}(\Phi)}\right)^{2}}\,.\label{renor}
\end{eqnarray}
In terms of the canonically normalized field we come up with the standard fashion
\begin{eqnarray}
\mathcal{S}_{\rm CI,E} =\int d^{4}x \sqrt{-g}\Big[\frac{1}{2} M_{\rm P}^2 R - \frac{1}{2}g^{\mu \nu} \partial_{\mu} \chi \partial_{\nu} \chi- {\cal U}({\chi}) \Big]\quad{\rm with}\quad {\cal U}({\chi})\equiv \Omega ^{-4} V({\Phi})\,. \label{EinsteinRe}
\end{eqnarray}
In this work, we will examine the dynamics in the Einstein frame, and therefore express the slow-roll parameters in terms of ${\cal U}$ and $\chi$:
\begin{eqnarray}
\epsilon &=& \frac{M_{\rm P}^2}{2}\left(\frac{d\,{\cal U}/d\chi}{{\cal U}}\right)^{2}=\frac{M_{\rm P}^2}{2}\left(\frac{{\cal U}'}{{\cal U}}\frac{1}{\chi'}\right)^{2}\,,\label{epsi}\\
\eta &=& M_{\rm P}^2 \frac{d^{2}\,{\cal U}/d\chi^{2}}{{\cal U}}=M_{\rm P}^2 \frac{{\cal U}''\chi'-{\cal U}'\chi''}{{\cal U}\chi'^{3}}\,, \label{eta}
\end{eqnarray}
where \lq\lq\,$\prime$\,\rq\rq\,denotes derivative with respect to the field $\Phi$. Slow-roll inflation ends when $\epsilon=1$, with the corresponding field value $\Phi_{end}$. The horizon exist when the field value equals $\Phi_{ini}$ which is determined by the number of e-foldings ${\cal N}$. This parameter is given by
\begin{eqnarray}
{\cal N}= \frac{1}{M_{\rm P}^2}\int^{\chi_{ini}}_{\chi_{end}}\frac{{\cal U}}{d\,{\cal U}/d\chi}d\chi=\frac{1}{M_{\rm P}^2}\int^{\Phi_{ini}}_{\Phi_{end}}\frac{{\cal U}\chi'^{2}}{{\cal U}'}d\Phi\,. \label{efold}
\end{eqnarray}
Here we express the number of e-foldings for the change of the field $\chi$ (or equivalently $\Phi$) from $\chi_{end}$ to $\chi_{ini}$ and will use the above mathematical framework to examine the composite-field models in Sec.(\ref{model}).

\section{Inflationary Observables}
\label{observable}

In this section, we will briefly review the study of the scalar and tensor perturbations. This leads to obtaining the required inflationary predictions, e.g. the spectral index and tensor-to-scalar ratio. Our starting point here is the action given in Eq.(\ref{EinsteinRe}). We will derive the parameters by closely following the work presented in \cite{Kobayashi:2011nu} for \lq\lq Generalised G-inflation\rq\rq\,using the unitary gauge. In the unitary gauge $\chi=\chi(t)$, it is kind of tradition to begin with the perturbed metric as \cite{Arnowitt:1962hi}
\begin{eqnarray}
ds^{2} =-N^{2}dt^{2}+\gamma_{ij}\Big(dx^{i}+N^{i}dt\Big)\Big(dx^{j}+N^{j}dt\Big)\,, \label{pert1}
\end{eqnarray}
where $N$ is a lab function, and $N^{i}$ is the shift vector which are respectively defined as:
\begin{eqnarray}
N=1+\alpha\,,\,\,\,N_{i}=\partial_{i}\beta\,,\,\,\,\gamma_{ij}=a^{2}(t)e^{2\zeta}\Big(\delta_{ij}+h_{ij}+\frac{1}{2}h_{ik}h_{kj}\Big)\,, \label{pert}
\end{eqnarray}
where $\alpha,\,\beta$ and $\zeta$ scalar perturbations and $h_{ij}$ is a tensor perturbation satisfying the condition $h_{ii}=0=h_{ij,j}$. Physically, the lapse function represents the rate of flow of proper time with respect to the coordinate time. Substituting the perturbed line-element given in Eq.(\ref{pert1}) into the action (\ref{EinsteinRe}) and expanding it to the second order, we obtain \cite{Kobayashi:2011nu}
\begin{eqnarray}
{\cal S}^{(2)}_{T} =\frac{1}{8}\int dt d^{3}{\bf x}\,a^{3}\left[\dot{h}^{2}_{ij}-\frac{1}{a^{2}}\Big(\vec{\nabla} h_{ij}\Big)^{2}\right]\,. \label{pertT}
\end{eqnarray}
Note that with the definition in \cite{Kobayashi:2011nu} in this case the sound speed equals unity, $c_{T}=1$. The tensor perturbations can be canonically renormalised via the following re-definitions:
\begin{eqnarray}
d\tau := \frac{1}{a}dt\,,\,\,z := \frac{a}{2}\,,\,\,v_{ij} := zh_{ij}\,. \label{define}
\end{eqnarray}
Plugging these new parameters, the above quadratic action becomes
\begin{eqnarray}
{\cal S}^{(2)}_{T} =\frac{1}{2}\int dt d^{3}{\bf x}\left[\Big(v'_{ij}\Big)^{2}-\Big(\vec{\nabla} v_{ij}\Big)^{2}+\frac{z''}{z}v_{ij}^{2}\right]\,, \label{pertT1}
\end{eqnarray}
where a prime represents derivative with respective to the conformal time $\tau$. It is common to transform the action and re-express the resulting one in terms of the Fourier modes, and one can basically show that the evolution equation from the action in Eq.\,(\ref{pertT1}) is given by
\begin{eqnarray}
v''_{ij}-\Big(k^{2}-\frac{z''}{z}\Big)v_{ij}=0\,. \label{evolu}
\end{eqnarray}
The normalized solution to this perturbation equation can be conventionally written in terms of the Hankel function so that we obtain the power spectrum of the primordial tensor perturbation at the horizon exits $c_{T}k|\tau|=1$ \cite{Kobayashi:2011nu} as
\begin{eqnarray}
{\cal P}_{T}=8\gamma_{T}\frac{H^{2}}{4\pi}\Big|_{c_{T}k|\tau|=1}\,, \label{pT}
\end{eqnarray}
where
\begin{eqnarray}
\gamma_{T} :=2^{2\nu_{T}-3}\Big|\Gamma(\nu_{T})/\Gamma(3/2)\Big|^{2}\Big(1-\epsilon\Big)\,,\,\,\nu_{T} \simeq 3/2+\epsilon+{\cal O}(\epsilon^{2})\,, \label{defi}
\end{eqnarray}
and the tensor spectral tilt is given by $n_{T}=3-2\nu_{T}=-2\epsilon+{\cal O}(\epsilon^{2})$. Now we turn to considering the scalar fluctuations. This can be achieved by setting $h_{ij}=0$ in the perturbed line-element mentioned in Eq.\,(\ref{pert}). Here we proceed in the same manner as for the tensor perturbations. For the scalar perturbations, we arrive with the second order action \cite{Kobayashi:2011nu}
\begin{eqnarray}
{\cal S}^{(2)}_{S} = \int dt d^{3}{\bf x}\,a^{3}\left[\epsilon\dot{\zeta}^{2}-\frac{\epsilon}{a^{2}}\Big(\vec{\nabla} \zeta\Big)^{2}\right]\,, \label{spertT}
\end{eqnarray}
where $\epsilon$ is already given in Eq.\,(\ref{epsi}), and the sound speed is also unity, $c_{S}=1$, in this case. The above action can be canonically renormalised via the following re-definitions:
\begin{eqnarray}
d\tau := \frac{1}{a}dt\,,\,\,z := \sqrt{2\epsilon}a\,,\,\,u := z\zeta\,. \label{sdefine}
\end{eqnarray}
Plugging these new parameters, the above quadratic action becomes
\begin{eqnarray}
{\cal S}^{(2)}_{S} =\frac{1}{2}\int dt d^{3}{\bf x}\left[\Big(u'\Big)^{2}-\Big(\vec{\nabla} u\Big)^{2}+\frac{z''}{z}u^{2}\right]\,, \label{spertT1}
\end{eqnarray}
where a prime represents derivative with respective to the conformal time $\tau$. It is common to transform the action and re-express the resulting one in terms of the Fourier modes, and one can basically show that the evolution equation from the action (\ref{spertT1}) is given by
\begin{eqnarray}
u''-\Big(k^{2}-\frac{z''}{z}\Big)u=0\,. \label{sevolu}
\end{eqnarray}
The normalized solution to this perturbation equation can be conventionally written in terms of the Hankel function so that we obtain the power spectrum of the primordial scalar perturbation at the horizon exits $c_{S}k|\tau|=1$ \cite{Kobayashi:2011nu} as
\begin{eqnarray}
{\cal P}_{S}=\frac{\gamma_{S}}{2}\frac{H^{2}}{4\pi^2\epsilon}\Big|_{c_{S}k|\tau|=1}\,, \label{spT}
\end{eqnarray}
where
\begin{eqnarray}
\gamma_{S} :=2^{2\nu_{S}-3}\Big|\Gamma(\nu_{S})/\Gamma(3/2)\Big|^{2}\Big(1-\epsilon\Big)\,,\,\,\nu_{S} \simeq 3/2+3\epsilon-\eta-\epsilon\,\eta+{\cal O}(\epsilon^2,\eta^2)\,, \label{sdefi}
\end{eqnarray}
and the spectral index is given by $n_{s}-1=3-2\nu_{S}=-6\epsilon+2\eta+{\cal O}(\epsilon^2,\eta^2)$. Moreover, the tensor-to-scalar ratio is defined by
\begin{eqnarray}
r\equiv \frac{{\cal P}_{T}}{{\cal P}_{S}}=16\epsilon\,. \label{t2s}
\end{eqnarray}
Recently, the Planck satellite data showed that the spectral index $n_{s}$ of curvature perturbations is constrained to be $n_{s} = 0.9603 \pm 0.0073$ ($68\%$ CL) and ruled out the exact scale-invariance ($n_{s} =1$) at more than $5\sigma$ confident level (CL), whilst the tensor-to-scalar ratio $r$ is bounded to be $r < 0.11$ ($95\%$ CL). Most recently, the observed $B$-mode power spectrum provides the tensor-to-scalar ratio $r=0.20^{+0.07}_{-0.05}$ with $r=0$ disfavoured at $7.0\sigma$ CL \cite{Ade:2014xna}. These constraints are used to falsify the most popular and simple inflationary models.

\section{Predictions from Composite Inflation}
\label{model}

According to the composite paradigms considered in this work, it was potentially shown that the models of composite inflation nicely respect tree-level unitary for the scattering of the inflation field during inflation all the way to the Planck scale. In this section, we consider composite inflationary models that can be described by the primordial power spectrum observables consisting of the spectral index, $n_{s}$, and the tensor-to-scalar ratio, $r$, derived in the Einstein frame. In the following computations, we assume a large non-minimal coupling $\xi\gg 1$ and a large field approximation, i.e. $\varphi^{2}\gg M^{2}_{\rm P}/\xi$, is also implemented. For later convenience, we express the inflationary parameters in terms of the number of e-foldings..

$\bullet$\,\,{\it Minimal Composite Inflation} (MCI)

The first model of composite inflation we will examine is recently investigated in \cite{Channuie:2011rq}. According to this paradigm, they engaged the simplest models of technicolor passing precision tests well known as the minimal walking technicolor (MWT) theory \cite{Sannino:2004qp,Hong:2004td,Dietrich:2005wk,Dietrich:2005jn} with the standard (slow-roll) inflationary paradigm as a template for composite inflation. The inflaton field in this case is the lightest composite state emerging from a bilinear condensate of techni-quarks with $d=1$. With the large field approximation, the action in the Jordan frame reads \cite{Channuie:2011rq}
\begin{eqnarray}
{\cal S}_{\rm MCI,\,J}=\int d^{4}x\sqrt{-g}\left[\frac{1 +\xi\varphi^{2}}{2} R - \frac{1}{2}g^{\mu\nu} \partial_{\mu}\varphi\partial_{\nu}\varphi- \frac{\kappa}{4}\varphi^{4} \right]\,,
\end{eqnarray}
where $\kappa$ is a self coupling which is constrained by the underlying theory to be of the order of unity. For this model, we have
\begin{eqnarray}
F(\varphi) = 1 + \xi\varphi^{2} \quad {\rm and} \quad G = 1\,.
\label{fg-tc}
\end{eqnarray}
Here we can diagonalize the inflaton-gravity sector by performing the conformal transformation already discussed in the previous section. After taking the conformal transformation, the resulting action in the Einstein frame reads
\begin{eqnarray}
{\cal S}_{\rm MCI,\,E}=\int d^{4}x\sqrt{-g}\left[-\frac{M^{2}_{P}}{2} R - \frac{1}{2}g^{\mu\nu} \partial_{\mu}\chi\partial_{\nu}\chi- {\cal U}_{\rm MCI}(\chi)  \right]\,,
\end{eqnarray}
where for a large field approximation
\begin{eqnarray}
\frac{d\chi}{d\varphi}\simeq \frac{\sqrt{6}M_{\rm P}}{\varphi}\quad{\longrightarrow}\quad \chi \simeq \sqrt{6}M_{\rm P}\ln(\sqrt{\xi}\varphi/M_{\rm P})\,. \label{chi-tc}
\end{eqnarray}
This leads to the potential in the Einstein frame as
\begin{eqnarray}
{\cal U}_{\rm MCI}(\chi) \simeq \frac{\kappa M^{4}_{P}}{4\xi^{2}}\Big(1+e^{\frac{-2\chi}{\sqrt{6}M_{P}}}\Big)^{-2}\quad{\rm with}\quad \Omega(\chi)^{2}\simeq \exp\left(\frac{2\chi}{\sqrt{6}M_{\rm P}}\right)\,.
\label{u-tc}
\end{eqnarray}
With the large field approximation, we can derive the following slow-roll parameter in terms of the field $\varphi({\rm or}\,\chi)$ as
\begin{eqnarray}
\epsilon \simeq \frac{4M^{4}_{\rm P}}{3\xi^{2}\varphi^{4}}\left({\rm or}\,\frac{4}{3}\exp\left(-\frac{4\chi}{\sqrt{6}M_{\rm P}}\right)\right)\,.\label{eptc}
\end{eqnarray}
Inflation ends when $\epsilon=1$ such that
\begin{eqnarray}
\varphi_{\rm end}\simeq \frac{M_{\rm P}}{\sqrt{\xi}}\left({\rm or}\,\chi_{\rm end}\simeq 0.3 M_{\rm P}\right)\,.\label{eptc1}
\end{eqnarray}
In the large field limits, the number of e-foldings reads
\begin{eqnarray}
{\cal N} \simeq \frac{6}{8M^{2}_{\rm P}/\xi}\Big[\varphi^{2}_{\rm ini}-\varphi^{2}_{\rm end}\Big]\quad{\rm with}\quad \varphi_{\rm ini}\gg\varphi_{\rm end}\,.\label{foldtc}
\end{eqnarray}
Using eqs.(\ref{epsi})-(\ref{efold}), it is straightforward to express the inflationary predictions in terms of the number of e-foldings, and find the Einstein frame parameters:
\begin{eqnarray}
n_{s}=1-6\epsilon+2\eta\simeq1-\frac{2}{{\cal N}}\,,\,\,\,\,r=16\epsilon\simeq\frac{12}{{\cal N}^2}\,. \label{paraTC}
\end{eqnarray}
Here we kept to the lowest order in $1/\xi$ to derive the spectral index and the tensor-to-scalar ratio. It is worthy to note here that the higher corrections of $1/\xi$ and $1/{\cal N}$ are suppressed to these expressions. This model provides $n_{s}\subseteq [0.933,\,0.975]$ and $r\subseteq [0.002,\,0.013]$ for ${\cal N}\subseteq [30,\,80]$. Unfortunately, from Eq.~(\ref{paraTC}), the predictions are in tension with the recent BICEP2 observations since a large value of $r$ cannot be produced in this model, see detailed discussions in the last section.   

$\bullet$\,\,{\it Glueball Inflation} (GI)

In this section, we consider the work presented in \cite{Bezrukov:2011mv} as the model motivated by a pure Yang-Mills theory. In this case, the inflaton is the lightest glueball field with $d=4$. The reason why we want to continue analysing composite inflation by following such model resides from the fact that this theory features the archetype of any composite paradigm in flat space and consequently of models of composite inflation. The low-energy effective Lagrangian of the lightest glueball state can be found in \cite{Schechter:1980ak,Migdal:1982jp,Cornwall:1983zb}. 

It is worthy to note here that the theory we are using describes the ground state of pure Yang-Mills theory, and of course is not the simple $\phi^4$ theory. Another difference from the $\phi^4$ theory is that the form of the effective potential, before coupling to gravity, is completely fixed by the underlying gauge theory. The action in the Jordan frame in this case reads \cite{Bezrukov:2011mv}
\begin{eqnarray}
{\cal S}_{\rm GI,\,J} = \int d^{4}x\sqrt{-g}\Big[\frac{1 + \xi\varphi^{2}}{2} R - 16g^{\mu\nu} \partial_{\mu}\varphi\partial_{\nu}\varphi - 2\varphi^{4}\ln\left(\varphi/\Lambda\right)  \Big]\,, \label{varaction}
\end{eqnarray}
which yields
\begin{eqnarray}
F(\varphi) = 1 + \xi\,\varphi^{2} \quad {\rm and} \quad G = 16\,.
\label{fg-gbr}
\end{eqnarray}
Imposing the conformal transformation, the resulting action in the Einstein frame reads
\begin{eqnarray}
{\cal S}_{\rm GI,\,E}=\int d^{4}x\sqrt{-g}\left[-\frac{M^{2}_{P}}{2} R - \frac{1}{2}g^{\mu\nu} \partial_{\mu}\chi\partial_{\nu}\chi- {\cal U}_{\rm GI}(\varphi(\chi))  \right]\,, 
\end{eqnarray}
which leads to the potential in the Einstein frame for a large field approximation as
\begin{eqnarray}
{\cal U}_{\rm GI}(\varphi) \simeq \frac{2M^{4}_{P}}{\xi^{2}}\ln\Big(\varphi/\Lambda\Big)\quad{\rm with}\quad \frac{d\chi}{d\varphi}\simeq \frac{\sqrt{6}M_{\rm P}}{\varphi}\longrightarrow \chi\simeq \sqrt{6}M_{\rm P}\ln(\varphi/\Lambda)\,.
\label{v-gl}
\end{eqnarray}
Here we have left the explicit dependence on the field $\varphi$ possessing unity canonical dimension instead of using the canonically normalized new scalar field $\chi$. However, we can express the resulting potential in terms of the field $\chi$ by substituting $\ln(\varphi/\Lambda)=\chi/\sqrt{6}M_{\rm P}$ into the potential given in Eq.~(\ref{v-gl}). With the large field approximation, we can derive the following slow-roll parameter 
\begin{eqnarray}
\epsilon \simeq \frac{1}{12\ln(\varphi/\Lambda)^{2}}\,.\label{epgb}
\end{eqnarray}
Inflation ends when $\epsilon=1$ such that
\begin{eqnarray}
\ln(\varphi_{\rm end}/\Lambda)\simeq \frac{1}{2\sqrt{3}}\,.\label{ep1gb}
\end{eqnarray}
In the large field limits, the number of e-foldings reads
\begin{eqnarray}
{\cal N} \simeq 3\Big(\ln\(\varphi_{\rm ini}/\Lambda\)^{2} - \ln\(\varphi_{\rm end}/\Lambda\)^{2}\Big)\,.\label{foldgb}
\end{eqnarray}
Using eqs.(\ref{epsi})-(\ref{efold}), the inflationary predictions in terms of the number of e-foldings in the Einstein frame parameters read
\begin{eqnarray}
n_{s}=1-6\epsilon+2\eta\simeq1-\frac{3}{2{\cal N}}\,,\,\,\,\,r=16\epsilon\simeq\frac{4}{{\cal N}}\,, \label{paraBlue}
\end{eqnarray}
where we have assumed $\varphi_{\rm ini}\gg \varphi_{\rm end}$. We also have kept to the lowest order in $1/\xi$ to derive the spectral index and the tensor-to-scalar ratio and neglected the higher corrections of $1/\xi$ and $1/{\cal N}$ suppressed to these expressions. The predictions of this model from Eq.~(\ref{paraBlue}) provide $n_{s}\subseteq [0.950,\,0.981]$ and $r\subseteq [0.050,\,0.13]$ for ${\cal N}\subseteq [30,\,80]$. Concretely, more detailed discussions will be added in the last section.

$\bullet$\,\,{\it Super Yang-Mills Inflation} (SYMI)

The application of the supersymmetric version of a pure Yang-Mills (SYM) has been investigated in various cornerstones. Here we consider for inflation the bosonic part of the Veneziano-Yankielowicz effective theory. For inflationary model building, the gluino-ball state in the super Yang-Mills theory is assigned as the inflaton with $d=3$. The effective Lagrangian in supersymmetric gluodynamics was constructed in \cite{Veneziano:1982ah}. Having already stamped as the  model of composite inflation, the authors of \cite{Channuie:2012bv} took the scalar component part of the super-glueball and coupled it non-minimally to gravity. The action of the theory in term of the canonical-dimension field $\varphi$ reads \cite{Channuie:2012bv}
\begin{eqnarray}
{\cal S}_{\rm SYMI,\,J} = \int d^{4}x\sqrt{-g}\Big[\frac{1+N^{2}_{c}\xi\varphi^{2}}{2} R - \frac{9N^2_{c}}{\alpha}g^{\mu\nu} \partial_{\mu}\varphi\partial_{\nu}\varphi - 4\alpha N^{2}_{c}\varphi^{4}(\ln[\varphi/\Lambda])^{2} \Big]\,,
\end{eqnarray}
with $N_{c}$ a number of colours and $\alpha$ a $N_{c}$-independent quantity. With the action given above, we find
\begin{eqnarray}
F(\varphi) = 1 + N^{2}_{c}\xi\,\varphi^{2} \quad {\rm and} \quad G = \frac{9N^{2}_{c}}{\alpha}\,.
\label{fg-sgb}
\end{eqnarray}
However, the gravity-scalar coupled sector can basically be diagonalised by imposing the conformal transformation. We find the resulting action in the Einstein frame as
\begin{eqnarray}
{\cal S}_{\rm SYMI,\,E}=\int d^{4}x\sqrt{-g}\left[-\frac{M^{2}_{P}}{2} R - \frac{1}{2}g^{\mu\nu} \partial_{\mu}\chi\partial_{\nu}\chi- {\cal U}_{\rm SYMI}(\varphi(\chi))  \right]\,, 
\end{eqnarray}
where
\begin{eqnarray}
{\cal U}_{\rm SYMI}(\varphi) \simeq \frac{4\alpha}{N^{2}_{c}}\frac{M^{4}_{P}}{\xi^{2}}\ln\Big(\varphi/\Lambda\Big)^{2}\quad{\rm with}\quad \frac{d\chi}{d\varphi}\simeq \frac{\sqrt{6}M_{\rm P}}{\varphi}\longrightarrow \chi\simeq \sqrt{6}M_{\rm P}\ln(\varphi/\Lambda)\,, \label{Ymp}
\label{v-ymi}
\end{eqnarray}
where we have also left the explicit dependence on the field $\varphi$ instead of using the canonically normalized new field $\chi$. With the large field approximation, we can derive the following slow-roll parameter 
\begin{eqnarray}
\epsilon \simeq \frac{1}{3\ln(\varphi/\Lambda)^{2}}\,.\label{epsgb}
\end{eqnarray}
Inflation ends when $\epsilon=1$ such that
\begin{eqnarray}
 \ln(\varphi_{\rm end}/\Lambda)\simeq \frac{1}{\sqrt{3}}\,.\label{epsg1b}
\end{eqnarray}
Performing the similar approximations to those of the previous subsection, the number of e-foldings for this model in the large $\xi$ limit reads
\begin{eqnarray}
{\cal N} \simeq \frac{3}{2}\Big(\ln\(\varphi_{\rm ini}/\Lambda\)^{2} - \ln\(\varphi_{\rm end}/\Lambda\)^{2}\Big)\,.
\label{efold-sgb-l}
\end{eqnarray}
Using eqs.(\ref{epsi})-(\ref{efold}), the inflationary predictions, i.e. $n_{s},\,r$, in terms of the number of e-foldings in the Einstein frame read
\begin{eqnarray}
n_{s}=1-6\epsilon+2\eta\simeq1-\frac{2}{{\cal N}}\,,\,\,\,\,r=16\epsilon\simeq\frac{8}{{\cal N}}\,. \label{parasusyym}
\end{eqnarray}
In order to derive these parameters, we have kept to the lowest order in $1/\xi$ and neglect the higher corrections of $1/\xi$ and $1/{\cal N}$. We find for this model from Eq.~(\ref{parasusyym}) that $n_{s}\subseteq [0.933,\,0.975]$ and $r\subseteq [0.100,\,0.267]$ for ${\cal N}\subseteq [30,\,80]$. In the last section, we will summarize our findings for this model.

$\bullet$\,\,{\it Orientifold Inflation} (OI)

The authors of \cite{Channuie:2012bv} examined the supersymmetric low-energy effective action to study inflation driven by the gauge dynamics of SU(N) gauge theories adding one Dirac fermion in either the two-index antisymmetric or symmetric representation of the gauge group. Such theories are known as orientifold theories. Here the gluino field of supersymmetric gluodynamics is replaced by two Weyl fields which can be formed as one Dirac spinor. For investigating the inflationary scenario, the orientifold sector non-minimally coupled to gravity in the Jordan frame action is implemented in \cite{Channuie:2012bv}. For this investigation, we write the action by using the real part of the field $\varphi$ as \cite{Channuie:2012bv}
\begin{eqnarray}
{\cal S}_{\rm OI,\,J}\supset \int d^{4}x\sqrt{-g}\left[-\frac{M^{2}_{P}+N^{2}_{c}\xi\varphi^{2}}{2} R - \frac{9F(N_{c})}{\alpha}g^{\mu\nu} \partial_{\mu}\varphi\partial_{\nu}\varphi- 4\alpha F(N_{c})\varphi^{4}\left(\ln(\varphi/\Lambda)^{2} -\gamma\right)  \right]\,, 
\end{eqnarray}
where $F(N_{c})=N^{2}_{c}(1+{\cal O}(1/N_{c}))$, $\gamma=1/9N_{c}+{\cal O}(1/N^{2}_{c})$ and hereafter we will keep only leading order in $1/N_{c}$. With the action given above, we find
\begin{eqnarray}
F(\varphi) = 1 + N^{2}_{c}\xi\,\varphi^{2} \quad {\rm and} \quad G = \frac{9F(N_{c})}{\alpha}\,.
\label{fg-sgb}
\end{eqnarray}
As usually proceeded in the standard fashion, we impose the conformal transformation and then find the resulting action in the Einstein frame    
\begin{eqnarray}
{\cal S}_{\rm OI,\,E}\supset\int d^{4}x\sqrt{-g}\left[-\frac{M^{2}_{P}}{2} R - \frac{1}{2}g^{\mu\nu} \partial_{\mu}\chi\partial_{\nu}\chi- {\cal U}_{\rm OI}(\varphi(\chi))  \right]\,, 
\end{eqnarray}
where
\begin{eqnarray}
{\cal U}_{\rm OI}(\varphi) \simeq \frac{4\alpha F(N_{c})}{N^{4}_{c}}\frac{M^{4}_{P}}{\xi^{2}}\Big[\ln\Big(\varphi/\Lambda\Big)^{2}-\gamma\Big]\quad{\rm with}\quad \frac{d\chi}{d\varphi}\simeq \frac{\sqrt{6}M_{\rm P}}{\varphi}\longrightarrow \chi\simeq \sqrt{6}M_{\rm P}\ln(\varphi/\Lambda)\,,
\label{v-oii}
\end{eqnarray}
with $N_{c}$ being a number of colours. Note that at large-$N_{c}$ the theory features certain super Yang-Mills properties, i.e. $F(N_{c})\rightarrow N^{2}_{c}$. With this limit, the transformed potential reduces to that of (\ref{Ymp}). With the large field limit, we can derive the following slow-roll parameter 
\begin{eqnarray}
\epsilon \simeq \epsilon_{\rm SYMI}\Big[1+\frac{2\gamma}{9}\ln(\varphi/\Lambda)^{2}\Big]\quad{\rm with}\quad\epsilon_{\rm SYMI}\simeq \frac{1}{3\ln(\varphi/\Lambda)^{2}}\,.\label{epoi}
\end{eqnarray}
Inflation ends when $\epsilon=1$ such that
\begin{eqnarray}
 \varphi^{3}_{\rm end}\simeq \varphi^{3}_{\rm SYMI,\,end}\Big[1+\frac{\gamma}{9\sqrt{3}}\Big]\,, \label{epsg1oi}
\end{eqnarray}
where $\varphi_{\rm SYMI,\,end}$ can be directly obtained from Eq.~(\ref{epsg1b}). According to this model, the number of e-foldings in the large $\xi$ limit is approximately given by
\begin{eqnarray}
{\cal N} \simeq \left[\frac{9\ln(\varphi/\Lambda)}{2\alpha\xi}\left(1-\frac{2\ln(3\ln(\varphi/\Lambda))}{81\ln(\varphi/\Lambda)^{2}}\right)\right]^{\varphi_{\rm ini}}_{\varphi_{\rm end}}\,.
\label{efold-oib-l}
\end{eqnarray}
Using eqs.(\ref{epsi})-(\ref{efold}), the predictions in the Einstein frame for $n_{s}$ and $r$ of this model can consequently written in terms of the number of e-foldings as
\begin{eqnarray}
n_{s}=1-6\epsilon+2\eta\simeq 1-\frac{2}{{\cal N}}\left(1+\frac{9\gamma}{2{\cal N}}\right)\,,\,\,\,\,r=16\epsilon\simeq \frac{8}{{\cal N}}\left(1+\frac{3\gamma}{{\cal N}}\right)\,. \label{paraorient}
\end{eqnarray}
Notice that for large $N_{c}$ the observables given above features the Super Yang-Mills inflation since $\gamma\rightarrow 0$.
\begin{figure}
\begin{center}
\includegraphics[width=4.5in]{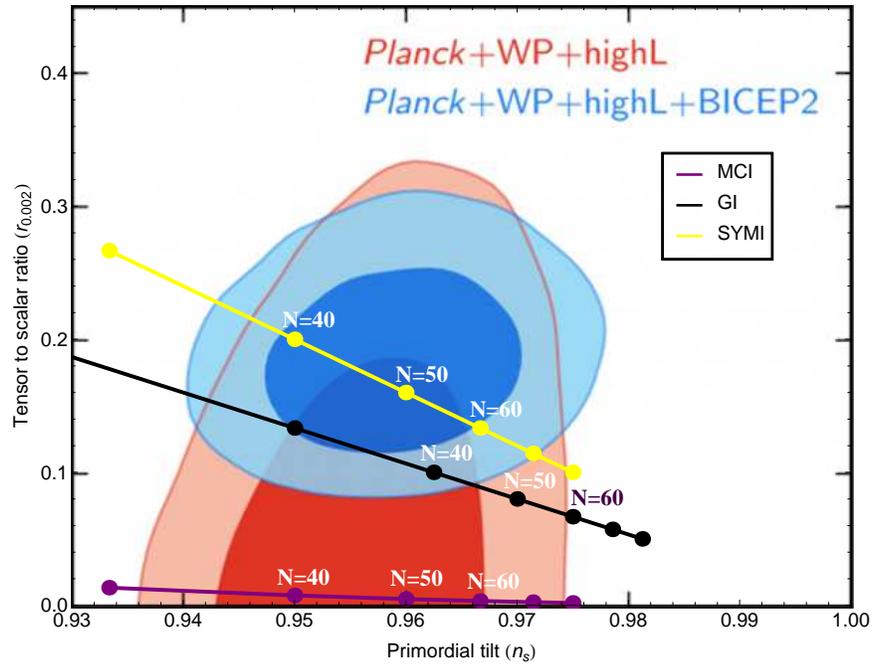} 
\end{center}
\caption{The contours show the resulting 68$\%$ and 95$\%$ confidence regions for the tensor-to-scalar ratio $r$ and the scalar spectral index $n_{s}$. The red contours are for the Planck\,data combination, which for this model extension gives a 95$\%$ bound $r < 0.26$ (Planck Collaboration XVI 2013 \cite{Ade:2013uln}). The blue contours represent the BICEP2 constraint on $r$ shown in the centre. The dot-line joints show the results of composite models examined in this work (MCI,\,GI,\,SYMI) assuming the number of e-foldings ${\cal N}$ to the end of inflation ranges in the interval ${\cal N} \subseteq [30,80]$.} \label{f1}
\end{figure}

\section{Summary}
\label{con}
In this last section we summarise our findings by which we confront the predictions of composite inflationary models with Planck and also very recent BICEP2 data and then plot our results on the constraint contours illustrated in Fig.\,(\ref{f1}). Moreover, recent study in the Jordan frame the primordial spectral index and tensor-to-scalar ratio predicted by composite models has been discussed in \cite{Channuie:2013lla}. For this work, the following is our remarks.
 
The first model of composite inflation (MCI) is inspired by models of dynamical electroweak symmetry breaking with the Higgs-like potential. Here the Higgs sector of the SM is replaced by a new underlying four-dimensional gauge dynamics free from fundamental scalars and the Higgs field is therefore composite state. We discover for model that the predictions can satisfy the $1\sigma$ CL of the Planck data for $35 \lesssim {\cal N} \lesssim 60$. Moreover, the model predictions can be consistent with the Planck data up to the $2\sigma$ CL if we choose $33 \lesssim {\cal N} \lesssim 80$. However, we find that the MCI model is in tension with the recent BICEP2 data. This is so since the model predictions yield quite small values of both $r$ and $n_{s}$ if we consider large $\xi$ limit. For a small $\xi$ case, there have been discussed in \cite{Channuie:2013lla}. In order to satisfy up to the $2\sigma$ CL of Planck data, we discover the number of e-foldings ${\cal N}$ should not be greater than $80$.

The underlying theory of the second paradigm (GI) features the archetype of any composite paradigm in flat space and consequently of models of composite inflation. The observables predicted by GI model lie inside the $1\sigma$ CL of the Planck for $30 \lesssim{\cal N} \lesssim 45$, whilst for $25 \lesssim {\cal N} \lesssim 60$ this model can be consistent with the $2\sigma$ region of the Planck contours. Interestingly, we discover for this model that the predictions can satisfy the $1\sigma$ CL of the BICEP2 data for $30 \lesssim {\cal N} \lesssim 35$, and up to $2\sigma$ CL if we choose $25 \lesssim {\cal N} \lesssim 45$. In order to satisfy upto the $2\sigma$ CL of Planck and BICEP2 data, we discover the number of e-foldings ${\cal N}$ should not be greater than $60$ and $45$, respectively.

Finally, the SYMI model features the supersymmetric version of a pure Yang-Mills properties. The model provide an interesting class of inflationary model. Apparently, the SYMI predictions are significantly consistent with the Planck and BICEP2 constraints. We find that in order to satisfy at the $1\sigma$ CL of the BICEP2 data, the model prefers $40\lesssim {\cal N}\lesssim60$. The model can be consistent with the 68$\%$ CL for both Planck and simultaneously BICEP2 data for $45\lesssim {\cal N}\lesssim60$, and with $95\%$ CL for $37\lesssim {\cal N}\lesssim70$. For the OI model, the present of $\gamma$ parameter yields small change of the predictions. However, the results from the OI model coincide with those of the SYMI for a large number of colours. In order to satisfy the $1\sigma$ and $2\sigma$ CL of the BICEP2 data, we discover the number of e-foldings ${\cal N}$ should not be greater than $60$ and $70$, respectively.

Another crucial consequence for the model of inflation is the (p)reheating mechanism. We anticipate to investigate this mechanism by following closely references \cite{Tsujikawa:1999iv,Watanabe:2006ku}. Rather interestingly, for instance, the author of \cite{Ratra:1991bn} studied the consequent of inflation as seed of the present intergalactic magnetic field. However, the author claimed that the results after making a number of simplifying approximations should be considered to be preliminary. Therefore, it is very interesting to study the mechanism for generating an intergalactic magnetic field based on the composite inflationary manners.

\noindent \underline{Acknowledgment}~:~ The author thanks Arunee Shunava and Worakrit Thida for assisted graphic works. Remote conversation with the CP$^3$-Origins's Team, F. Sannino, J. Joergensen and O. Svendsen, is also acknowledged. This work is fully supported by the \lq\lq Research Fund for DPST Graduate with First Placement\rq\rq\,under Grant No. 033/2557. 


\bibliographystyle{unsrt}

\end{document}